\documentclass{webofc}

\usepackage[varg]{txfonts}   
\usepackage{hyperref}
\usepackage{url}
\usepackage{bm}
\hypersetup{colorlinks=true,citecolor=blue,urlcolor=blue,linkcolor=blue}
\usepackage{calc,atbegshi,picture}
\newcommand{\placepreprint}{%
\AtBeginShipoutFirst{%
\put(
\textwidth+\oddsidemargin-\widthof{ MIT-CTP/5724},
-2pt-\topmargin-\heightof{ MIT-CTP/5724}
){\normalfont MIT-CTP/5724}
}
}
\begin{document}
\placepreprint
\title{Connecting Pre-Thermal and Hydrodynamizing Attractors With Adiabatic Hydrodynamization}
%
%

\author{\firstname{Krishna} \lastname{Rajagopal}\inst{1}\fnsep\thanks{\email{krishna@mit.edu}} \and
        \firstname{Bruno} \lastname{Scheihing-Hitschfeld}\inst{1,2}\fnsep\thanks{\email{bscheihi@kitp.ucsb.edu}} \and
        \firstname{Rachel} \lastname{Steinhorst}\inst{1}\fnsep\thanks{\email{rstein99@mit.edu;} speaker at Hard Probes 2024}
}

\institute{Center for Theoretical Physics, Massachusetts Institute of Technology, Cambridge, MA 02139,
USA 
\and
           Kavli Institute for Theoretical Physics, University of California, Santa Barbara, California 93106,
USA
          }

\abstract{The far-from equilibrium dynamics of the pre-hydrodynamic quark-gluon plasma (QGP) formed in heavy ion collisions can be characterized by distinct stages, during each of which the system loses some memory of its initial condition, until only the hydrodynamic modes remain. However, even though it has been repeatedly observed, finding intuitive physical explanations of how and why attractor behavior occurs has remained a challenge. The Adiabatic Hydrodynamization (AH) framework provides exactly such an explanation, showing that the attractor solution can be thought of as the ground state of an analog to quantum mechanical adiabatic evolution, provided we identify appropriate coordinate rescalings. Using the example of a simplified QCD kinetic theory in the small-angle scattering limit, we show how AH can explain both the early pre-hydrodynamic attractor and the later hydrodynamizing attractor in a longitudinally expanding gluon gas in a unified framework. By doing this, we provide a unified description of, and intuition for, all the stages of what in QCD would be bottom-up thermalization, starting from a pre-hydrodynamic attractor and ending with hydrodynamization.
}
\maketitle
The early success of hydrodynamics in describing relativistic heavy ion collision data from RHIC (and later from the LHC) was astonishing in part because it required rapid equilibration of the quark-gluon plasma (QGP). In the time since this was realized, the feasibility of such rapid equilibration has been demonstrated both in strongly-coupled and weakly-coupled descriptions.

Indeed, in weakly-coupled descriptions it is well-established that the gas of far-from-equilibrium gluons undergoes a significant loss of degrees of freedom often long before hydrodynamics~\cite{Kurkela:2019set}, falling onto a seemingly universal non-thermal attractor solution. Though the presence of these attractors is known, physical intuition for why/how the system should be attracted to successive attractors has been lacking. The relationship between these attractors is also unclear: one could imagine scenarios in which the hydrodynamic mode shares no overlap with the non-thermal attractor mode, or in which the hydrodynamic mode is simply the late-time limit of a single time-dependent attractor. The Adiabatic Hydrodynamization (AH) framework~\cite{Brewer:2019oha} is a natural candidate for answering these questions, and we have used it to show that neither such scenario is realized. 

Using the Adiabatic Hydrodynamization approach for a simplified version of QCD kinetic theory, we showed in our recent paper~\cite{Rajagopal:2024lou} that the pre-thermal attractor mode is in reality a ``band" of nearly degenerate transverse modes, with the system having ``forgotten'' all aspects of its initial conditions that were originally encoded in modes from higher bands. Furthermore, we showed that when the hydrodynamic mode later emerges, it is formed exclusively as a single linear combination from the modes in the lowest band, with no mixing from other bands. Hydrodynamization is described via the energies of all but one states from the lowest band rising, leaving a lone mode that evolves adiabatically as it hydrodynamizes. Furthermore, the adiabatic picture provides an intuitive physical motivation for why the gluon distribution function should be attracted to these attractors: the pre-hydrodynamic and hydrodynamizing attractors are each, during their epoch, the effective time-dependent ground state of the system's evolution, with the occupation of higher modes decaying exponentially. We review these developments in what follows.

\section{Scaling, Attractors, and Adiabaticity}
Scaling is a manifestation of attractor behavior which has been observed in many pre-hydrodynamic descriptions. In the case of classical-statistical simulations and kinetic theory, the latter of which is the focus of our study, one can say that the distribution function is ``scaling'' if it takes the form \cite{Berges:2013eia}
\begin{align}\label{eq:scaling}
f_s(p_\perp,p_z,\tau) = \tau^{\alpha} w_s (p_\perp \tau^{-\beta},p_z \tau^{-\gamma})
\end{align}
If additionally the form of the underlying rescaled distribution function $w_s(\zeta,\xi)$ is universal, one can say that $w_s$ acts as an attractor. In such a case, we need only know $w_s$ and the scaling exponents $\alpha,\beta,\gamma$ in order to fully describe the system. With each stage of the traditional picture of bottom-up thermalization~\cite{Baier:2000sb}, one can roughly associate a set of scaling exponents. The best known is the ``BMSS" fixed point, $\alpha=-2/3,\,\beta=0,\,\gamma=-1/3$, and it is furthermore known that $f$ is ``prescaling" in the approach to the BMSS fixed point~\cite{Mazeliauskas:2018yef}, in the sense that it takes the form of Eqn.~\eqref{eq:scaling} for time-dependent  $\alpha,\beta,\gamma$. 

This inspires the definition of an effective Hamiltonian operator $H_{\rm eff}$ satisfying
\begin{align}
\label{eq:schrodinger}
H_{\rm eff}w = -\partial_y w
\end{align}
where $y\equiv \log(\tau/\tau_I)$ and $w$ is defined by
\begin{align}\label{eq:frescaled}
f(p_\perp,p_z,\tau) = A(\tau) w\left(\frac{p_\perp}{B(\tau)},\frac{p_z}{C(\tau)},\tau\right)\,.
\end{align}
This is natural in the sense that for $A,\,B\,,C$ satisfying
\begin{align}
\label{eq:scalingexponents}
\frac{\partial_y A}{A}=\alpha\,,\qquad \frac{\partial_y B}{B}=-\beta\,,\qquad \frac{\partial_y C}{C}=-\gamma\,,
\end{align}
when $f$ is prescaling, we will have $\partial_y w=0$. That is, $w_s$ is an instantaneous eigenstate of $H_{\rm eff}$ with ``effective energy" 0. As compared with the Schr\"{o}dinger equation (in addition to $H_{\rm eff}$ in general being nonlinear and non-Hermitian), the -1 in place of $i$ in Eqn.~(\ref{eq:schrodinger}) suggests that excited effective energy modes will tend to die off as $\sim e^{-\int^\tau \epsilon_i(\tau') \,d\tau'}$. This intuition will hold so long as $H_{\rm eff}$ evolves sufficiently slowly compared to the effective energy gap, analogously to the Adiabatic Theorem in Quantum Mechanics. 

If we are able to choose $A(\tau),\,B(\tau)\,,C(\tau)$ to simultaneously make $H_{\rm eff}$ adiabatic and satisfy Eqn.~(\ref{eq:scalingexponents}) in the prescaling regime, we have found an intuitive physical explanation for why $w$ is attracted to the attractor $w_s$: it is simply the ground state of an adiabatic time-dependent effective Hamiltonian $H_{\rm eff}$, revealed by solving the eigenvalue problem for $H_{\rm eff}$, as long as we make the appropriate choice of time-dependent coordinate rescaling. This explanation for prehydrodynamic and hydrodynamizing attractor behavior is provided by the Adiabatic Hydrodynamization (AH) framework, first proposed 
in Ref.~\cite{Brewer:2019oha} in the context of the hydrodynamizing attractor in RTA. AH was also demonstrated for the BMSS and ``dilute" prehydrodynamic scaling regimes using only the diffusive part of the small-angle elastic collision kernel~\cite{Brewer:2022vkq}. Only when incorporating the non-diffusive piece will the system be able to thermalize, as in Ref.~\cite{Rajagopal:2024lou} and in the results shown here. In this way, we use the AH framework to understand how the transition between early and late-time attractors occurs.

\section{Kinetic Theory and Setup}

Under the assumption of boost-invariance and homogeneity in the transverse spatial plane, the gluon distribution function $f$ will evolve according to a Boltzmann equation
\begin{align}
\frac{\partial f}{\partial \tau} -\frac{p_z}{\tau} \frac{\partial f}{\partial p_z} = -C[f]\,,
\end{align} 
where $C[f]$ is the collision kernel. As in Ref.~\cite{Brewer:2022vkq}, we will omit inelastic scatterings and take only the small-angle elastic collision kernel. Furthermore, due to numerical challenges presented by explicitly nonlinear components of the effective Hamiltonian, we omit the explicit Bose enhancement factor, meaning we have a Boltzmann rather than a Bose-Einstein solution to $C[f]$. Our collision kernel is given by
\begin{equation}\label{eq:collisionkernel}
  \mathcal{C}[f] = -\lambda_0 \ell_{\rm Cb} [f] \left[ I_a[f] \nabla_{\bf p}^2 f + I_b[f] \nabla_{\bf p} \cdot \left( \frac{\bf p}{p} f \right) \right]\, ,
\end{equation}
where $I_a[f]=\int_{\bm p}f(1+f)$, $I_b[f] = \int_{\bm p} f/p$, $\ell_{\rm Cb} = \ln (p_{\rm UV}/p_{\rm IR})$, and $\lambda_0=4\pi\alpha_s^2N_c^2$. We choose to retain the Bose factor in $I_a$ in order to see BMSS scaling.

The natural way to proceed is to perform coordinate rescalings as in Eqn.~\eqref{eq:frescaled} and solve for $H_{\rm eff}$. However, the $1/p$ appearing in the non-diffusive (proportional to $I_b$) part of the collision kernel~\eqref{eq:collisionkernel} means that $H_{\rm eff}$ cannot be decomposed into separable parts. This makes projecting $H_{\rm eff}$ onto a basis at each time step time consuming and in practice, infeasible. One can sidestep the issue by taking $p\approx p_\perp$, which is a good approximation at early times. We do this in Sec.~3.3.1 of Ref.~\cite{Rajagopal:2024lou}, and can verify in this way the conclusions of Ref.~\cite{Brewer:2022vkq} regarding the adiabaticity of early-time attractors with our generalized collision kernel.

However, because thermalization is accompanied by isotropization, taking $p\approx p_\perp$ eliminates our ability to describe the hydrodynamizing attractor. We can restore this ability by choosing a different parametrization for $f$,
\begin{align}
f(p,u,\tau) = A(\tau) w\left( \frac{p}{D(\tau)},u,r(\tau),\tau\right)\,,
\end{align}
where $u\equiv p_z/p$.
In practice $r(\tau)$ will be used to parametrize anisotropy, although it is not a true coordinate rescaling of $u$. Now, the connection between $w$ and $w_s$ in the BMSS or dilute regime is less clear, but thermal scaling is much more natural; we would expect that at late times
$\frac{\partial_y A}{A}=0$ and $\frac{\partial_y D}{D}=-\frac{1}{3}$,
and that $D$ is proportional to the effective temperature.

We then find that $H_{\rm eff}$ is
\begin{align}
    H_{\rm eff} &= \left(\frac{\partial_y A}{A}\right) + (\partial_y r) \partial_r - \left(\frac{\partial_y D}{D}\right) \chi \partial_\chi - u^2 \chi \partial_\chi - u(1-u^2) \partial_u \nonumber \\  
    & \quad - \tau \lambda_0 \ell_{\rm Cb} \frac{I_a}{D^2}  \left[ \frac{2}{\chi} \partial_\chi + \partial_\chi^2 + \frac{1}{\chi^2} \frac{\partial}{\partial u} \left( (1 - u^2) \frac{\partial f}{\partial u} \right) \right]  
    - \tau \lambda_0 \ell_{\rm Cb} \frac{I_b}{D} \left( \frac{2}{\chi} +  \partial_\chi \right) \, .
\end{align}
Now we need only make prudent choices for the time-dependent scalings $A(\tau),\,D(\tau),\,r(\tau)$, making sure that $H$ is slowly-evolving and in such a way that we capture the physical scaling behavior, and to solve for the evolution of the system by projecting it onto a basis. We choose $A(\tau)$ as suggested by number conservation,
$
\frac{\partial_y A}{A}=-3\frac{\partial_y D}{D}-1
$, 
and $D(\tau)$ such that it follows what at late times is the effective temperature,
$
    \frac{\partial_y D}{D} = 10 \left( 1 - D \left\langle \frac{2}{p} \right\rangle  \right) 
$. 
Guided by the knowledge that at early times $w_s\sim e^{-p_z^2/C(\tau)^2}$, and that at late times $w_s \sim e^{-p/T}$, we choose the right and left basis
\begin{align}
    \psi_{nl}^{(R)} = N_{nl} e^{-\chi} e^{-u^2 r^2/2} L_{n-1}^{(2)}(\chi) Q_l^{(R)}(u;r) \, , & & \psi_{nl}^{(L)} = N_{nl} L_{n-1}^{(2)}(\chi) Q_l^{(L)}(u;r) \, ,
\end{align}
where $Q_l^{(R/L)}(u;r)$ and $L_{n-1}^{(2)}(\chi)$ are orthogonal polynomials with respect to the integral measures defined by $e^{-u^2 r^2/2}$ and $\chi^2 e^{-\chi}$, respectively, and $N_{nl}$ are normalization constants. This basis has the quality that for small $r$, it mimics the early-time attractor, while for large $r$, it mimics the late-time attractor. Finally, we use our choice of $r(y)$ to attempt to maximize the extent to which $f$ is described by $\psi_{10}^{(R)}$, in particular by setting
$
\int_{\bf p} u^2 \left(\partial_y+H_{\rm eff}\right) \psi_{10}^{(R)} = 0
$
and solving for $\partial_y r$. For more details on the numerical implementation, see Ref.~\cite{Rajagopal:2024lou}.

\section{Results}

\begin{figure*}
    \centering
    \includegraphics[width=6cm,clip]{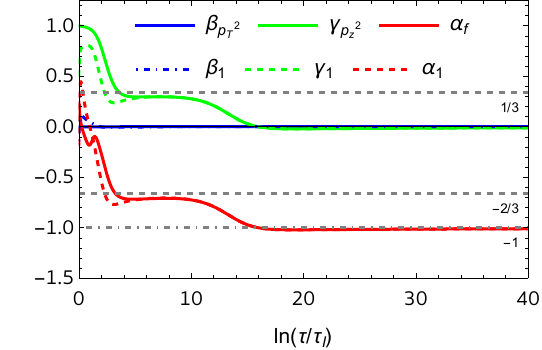}
    \hfill
    \includegraphics[width=6cm,clip]{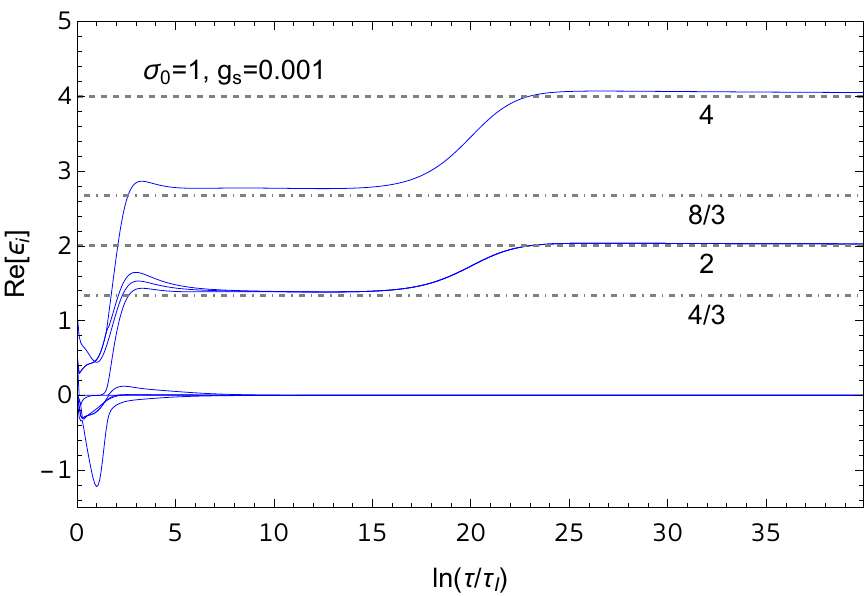}
    \caption{On the left, scaling exponents, and on the right, effective energy levels, each for $g_s=10^{-3}$. 
    Scaling exponents are as defined in Eqns.~\eqref{eq:physscaling} and~\eqref{eq:basisscaling}. The good agreement between the two scaling exponent definitions suggests our choice of basis and scaling describes the evolution of $f$ well. Both scaling regimes are characterized by well-separated bands of nearly degenerate modes, suggesting a partial loss of memory as the system falls onto the BMSS attractor. With a coupling this small, hydrodynamization does not occur.}  \label{fig:weakcoupling} 
\end{figure*}

\begin{figure*}
    \centering
    \includegraphics[width=6cm,clip]{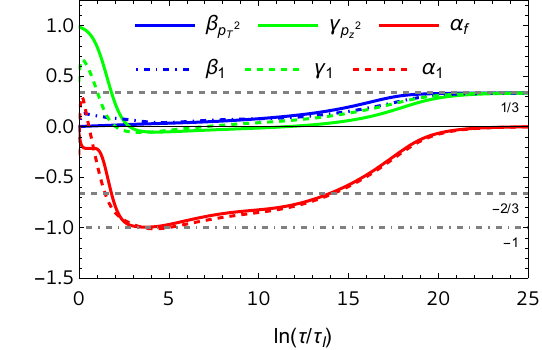}
    \hfill
    \includegraphics[width=6cm,clip]{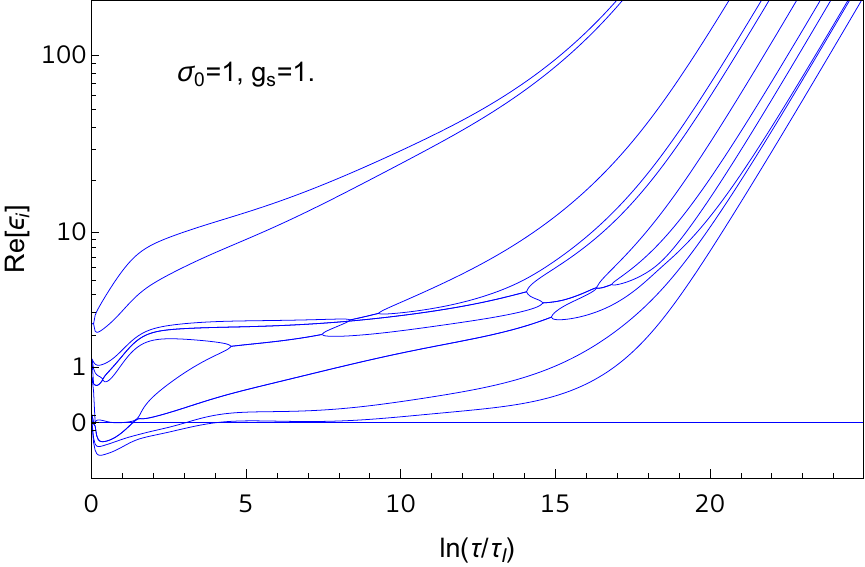}
    \caption{On the left, scaling exponents, and on the right, effective energy levels, each for $g_s=1$. 
    Scaling exponents are again as defined in Eqns.~\eqref{eq:physscaling} and~\eqref{eq:basisscaling}, and as in the weak coupling case agreement suggests a good time-dependent basis choice. Between times approximately $y=3$ and $y=15$, both scaling exponents and effective energy levels seem to resemble the dilute scaling regime, although not as cleanly as at weaker coupling. All but one of the effective energies which form the low-energy dilute ``band" lift off at late times $y \gtrsim 15$ to reveal a unique ground state, the hydrodynamizing attractor. As the gap appears, the scaling exponents approach their expected values in local thermal equilibrium -- the adiabatic evolution describes hydrodynamization.} \label{fig:strongcoupling} 
\end{figure*}

We choose the initial condition
\begin{equation} 
    f({\bf p}, \tau = \tau_I) = \frac{\sigma_0}{g_s^2} e^{- \sqrt{2} p/Q_s} e^{- r_I^2 u^2 /2 } Q_0^{(R)}(u;r_I)
\end{equation}
with $r_I = \sqrt{3}$, $\tau_I Q_s = 1$, and $\sigma_0 = 1$; and solve for its effective Hamiltonian evolution on our basis using 12 basis states for both very weak (Fig.~\ref{fig:weakcoupling}) and relatively stronger (Fig.~\ref{fig:strongcoupling}) choices of the coupling $g_s$. We define two sets of scaling exponents: First, to capture the scaling behavior of $f$, we define
\begin{align}\label{eq:physscaling}
    \beta_{\langle p_T^2\rangle} \equiv - \frac{1}{2} \partial_y \ln \langle p_\perp^2 \rangle \, ,  \quad
    \gamma_{\langle p_z^2\rangle} \equiv -\frac{1}{2} \partial_y \ln \langle p_z^2 \rangle \, , \quad
    \alpha_{\langle f \rangle} \equiv \partial_y \ln \langle f \rangle 
\end{align}
Calculating this quantity does not demonstrate that $f$ is scaling, but rather is meant to indicate in which scaling regime $f$ is as a function of time. Secondly, to analyze the aptness of our choices of basis and scalings, we define
\begin{align}\label{eq:basisscaling}
    \beta_1 \equiv -\frac12 \partial_y \ln \langle p_\perp^2 \rangle_1  \, , \quad
    \gamma_1 \equiv -\frac12 \partial_y \ln \langle p_z^2 \rangle_1 \, , \quad
    \alpha_1 \equiv \partial_y \ln \langle f_1 \rangle_1 \, ,
\end{align}
where the average $\langle \cdot \rangle_1$ denotes
$
    \langle X \rangle_1 \equiv \left(\int_{\bf p} X f_1 \right)/\int_{\bf p} f_1
$,
i.e.~this second set of exponents represent the scaling exponents of the first basis state only. If the two sets of scaling exponents agree, this suggests that $f$ is well-captured by our choice of time-dependent basis.

At weak coupling $g_s = 10^{-3}$, we see the expected BMSS and dilute scaling regimes, but do not see hydrodynamization (Fig.~\ref{fig:weakcoupling}). This is at least partially due to our omission of inelastic scatterings from the collision kernel, the inclusion of which would significantly hasten hydrodynamization. It is promising that the analytical prediction of~\cite{Brewer:2022vkq} for the effective energies using the diffusive part of the collision kernel only,
\begin{equation} \label{eq:energies}
    \epsilon_{nm} = 2n(\gamma-1)-2m\beta \;\;\;\;\; n,m=0,1,2,\dots\,.
\end{equation}
still clearly describes what is happening. However, the gap between transverse modes as parametrized by $m$ in Eqn.~\eqref{eq:energies}, is vanishingly small, and since $H_{\rm eff}$ is not perfectly time-independent in this case, the evolution is not adiabatic in the sense of having a unique ground state evolving independently from all other modes. Regardless, the concept of adiabaticity still applies to this problem if we make use of the separation of the ``bands" of effective eigenstates with differing $n$. Rather than a unique ground state which acts as an attractor, we instead have a set of modes which collectively act as an attractor surface. 

At a somewhat larger coupling $g_s = 1$, we finally see hydrodynamization (Fig.~\ref{fig:strongcoupling}). Hydrodynamization is preceded by a regime of dilute scaling: although the scaling exponents are not constant in this regime as seen at weaker coupling, the energies are still consistent with the dilute scaling expectation. Most interestingly, we can see from the effective energy levels that the hydrodynamizing attractor mode emerges as the unique ground state at late times from the ground state ``band" of the dilute regime, with no mixing from higher-energy modes. This means that the early- and late-time attractors are continuously connected, with memory loss occurring in two stages: a first stage in which all memory in excited longitudinal modes is forgotten (the BMSS and/or dilute stage), and a second in which all of the remaining memory is lost and the system falls onto the hydrodynamic mode. We have achieved a unified description that starts from the initial stage, proceeds through both attractors, and hydrodynamizes.

\section{Outlook}

We have demonstrated that the AH framework provides an intuitive physical explanation for attractor behavior, and in particular an explanation which unifies pre-thermal and thermal attractors. We have shown that the reduction of degrees of freedom in the out-of-equilibrium gluon gas we have studied occurs in distinct stages, and that the hydrodynamizing mode emerges from a low-energy band of modes which make up a pre-thermal attractor surface.

It would of course be interesting to extend this analysis to the full QCD EKT collsion kernel; one of the most important steps towards doing so is the incorporation of number non-conserving processes. We expect that this will dramatically decrease the time it takes for the kinetic theory to hydrodynamize, making the dynamics much more realistic. This is work in progress.
It will also be important to incorporate spatial gradients to have a clearer picture of precisely what kind of initial state information in the transverse spatial plane survives the process of memory loss in kinetic theory. Identifying the important modes to follow prior to hydrodynamization could provide an important tool to study the initial state in contexts where computational speed is important, for example, in Bayesian analyses of heavy ion collisions. This would add further to the impact of AH, beyond the physical intuition that it provides.

\section*{Acknowledgements}

This work is supported by the U.S.~Department of Energy, Office of Science, Office of Nuclear Physics under grant Contract Number DE-SC0011090\@. The research of BSH was supported in part by grant NSF PHY-2309135 to the Kavli Institute for Theoretical Physics (KITP). The research of BSH was supported in part by grant 994312 from the Simons Foundation.

\bibliography{main} 

\end{document}